\documentclass[a4paper,twocolumn,showpacs,superscriptaddress,floatfix,nofootinbib]{revtex4-1}
\renewcommand{\nat}{{Nature}}
\renewcommand{\apj}{{ApJ}}
\renewcommand{\prd}{{PRD}}
\renewcommand{\prc}{{PRC}}
\newcommand{\apjl}{{ApJL}}
\newcommand{\cqg}{{CQG}}
\newcommand{\mnras}{{MNRAS}}

\usepackage{latexsym}
\usepackage{amssymb}
\usepackage{amsfonts}
\usepackage{amsmath}
\usepackage{bm}
\usepackage{graphicx}   
\usepackage{color}
\usepackage{subfigure}
\usepackage{times}
\usepackage{units}
\usepackage{hyperref}
\usepackage{filecontents}

\begin{document}

\title{Short gamma-ray bursts from binary neutron star mergers: the time-reversal scenario}

\author{Riccardo Ciolfi}
\affiliation{Physics Department, University of Trento, Via Sommarive 14, 38123 Trento, Italy} 
\affiliation{INFN-TIFPA, Trento Institute for Fundamental Physics and Applications, Via Sommarive 14, 38123 Trento, Italy}

\author{Daniel M. Siegel}
\affiliation{Max Planck Institute for Gravitational Physics (Albert Einstein Institute), Am M\"uhlenberg 1, 14476 Potsdam-Golm, Germany}
        
\begin{abstract}        
After decades of observations the physical mechanisms that generate short gamma-ray bursts (SGRBs) still remain unclear. Observational evidence provides support to the idea that SGRBs originate
from the merger of compact binaries, consisting of two neutron stars (NSs) or a NS
and a black hole (BH).
Theoretical models and numerical simulations seem to converge to an
explanation in which the central engine of SGRBs is given by a spinning BH
surrounded by a hot accretion torus. Such a BH-torus system can be formed
in compact binary mergers and is able to launch a relativistic jet, which can then produce the SGRB.
This basic scenario, however, has recently been challenged by \textit{Swift} satellite observations, which have revealed long-lasting X-ray afterglows in association with a large fraction of SGRB events.
The long durations of these afterglows (from minutes to several hours) cannot
be explained by the $\sim$s accretion timescale of the torus onto the BH, and, instead,
suggest a long-lived NS as the persistent source of radiation.
Yet, if the merger results in a massive NS the conditions to
generate a relativistic jet and thus the prompt SGRB emission are hardly met.
Here we consider an alternative scenario that can reconcile the two aspects and
account for both the prompt and the X-ray afterglow emission. Implications for
future observations, multi-messenger astronomy and for constraining NS properties are discussed, as well as potential challenges for the model.
\end{abstract}

\maketitle

\section{Introduction}

\noindent Binary neutron star (BNS) mergers and neutron star-black hole (NS-BH) 
binary mergers represent the leading candidate scenarios to explain the phenomenology of 
short gamma-ray bursts (SGBRs) 
\cite{Paczynski86}.
Moreover, they are among the most promising sources of gravitational waves (GWs) that are 
likely to be detected in the near future with ground-based 
detectors such as advanced LIGO and Virgo \cite{Accadia2011}.
In the most studied scenario, the compact binary merger results in the formation 
of a spinning BH surrounded by a hot and thick accreting torus. During the 
short ($\lesssim 1$~s) accretion phase a relativistic jet can be launched via 
different mechanisms (involving neutrino annihilation and/or magnetic fields), finally 
producing the SGRB.
This picture is supported by recent numerical simulations (\cite{Rezzolla2011} 
and references therein), which have shown that a massive ($\sim 0.1$~M$_\odot$) 
accretion torus is naturally formed when the merger leads to the prompt formation 
of a spinning BH.

Recent observations by \textit{Swift} \cite{Gehrels2004} of long-lasting ($\sim 10^2-10^5$~s) X-ray afterglows in 
association with a large fraction of SGRBs, however, conflict with the 
above BH-torus model, since the short accretion timescale can hardly explain 
such a durable X-ray emission. 
As a possible alternative, the formation of a long-lived and highly-magnetized 
NS that continues injecting energy on much longer timescales via spin-down 
radiation can explain the observed X-ray afterglow durations and 
luminosities \cite{Zhang2001}.
Nevertheless, this so-called ``magnetar model'' does not provide an explanation 
for the generation of the prompt SGRB emission, which thus leads to an apparent dichotomy. 

Here we discuss the novel ``time-reversal'' scenario recently proposed in 
\cite{Ciolfi2015}, which can explain the prompt SGRB and the X-ray 
afterglow emission in a common phenomenology and, hence, solve the above 
dichotomy (for an alternative proposal, see \cite{RezzollaKumar}).
In particular, in Section~\ref{phenomenology} we briefly summarize the 
basic phenomenology and the results of \cite{Ciolfi2015}, while 
Sections~\ref{evidence} and \ref{constraints} focus on the specific predictions 
of the model and on the possibility of placing stringent constraints on 
NS properties. Finally, in Section~\ref{conclusion} we draw conclusions 
and discuss future work, also pointing out potential problems of the model. 

\begin{figure*}[t]
\centering 
\includegraphics[angle=0,width=0.32\textwidth]{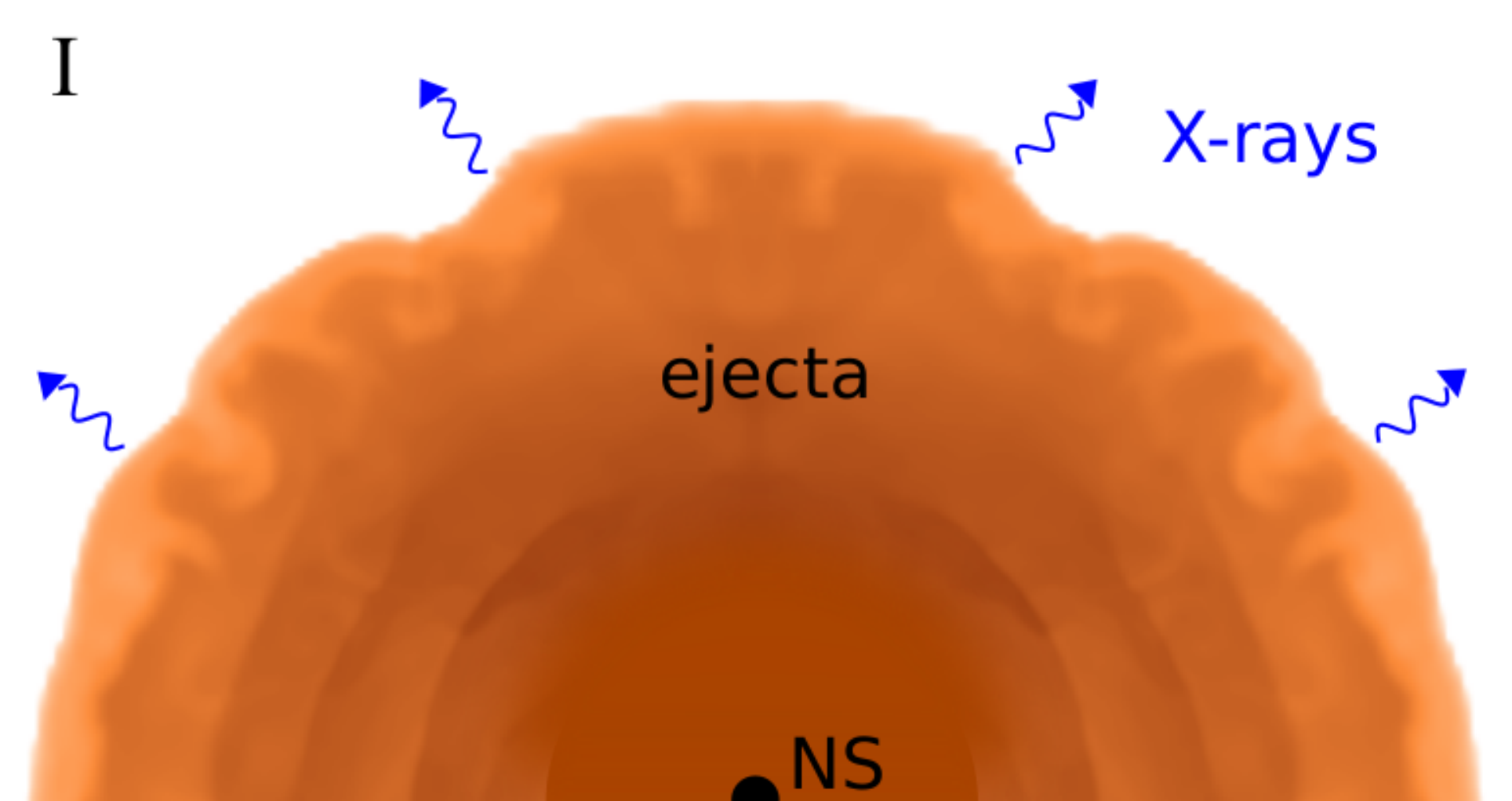}
\includegraphics[angle=0,width=0.32\textwidth]{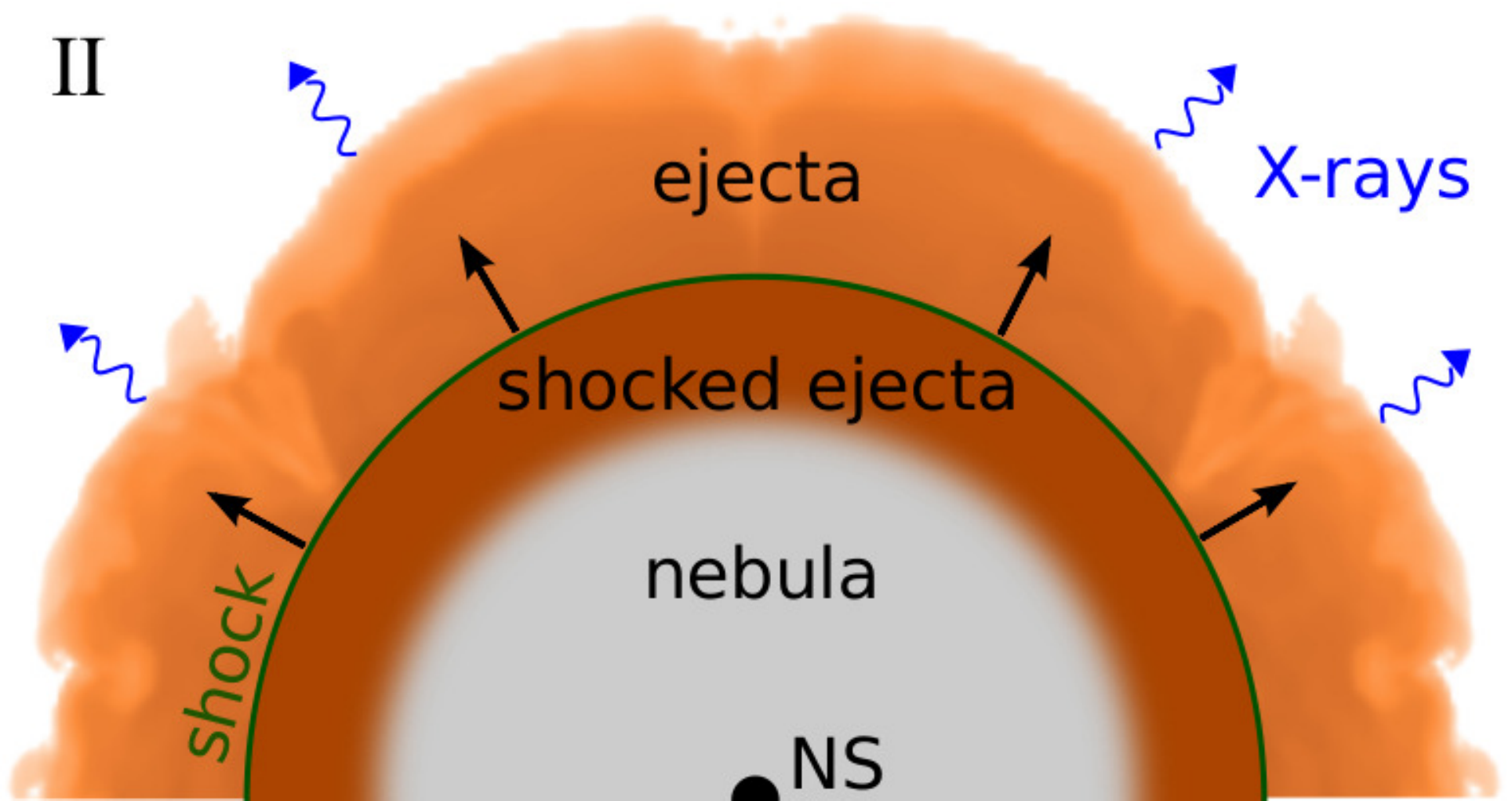}
\includegraphics[angle=0,width=0.32\textwidth]{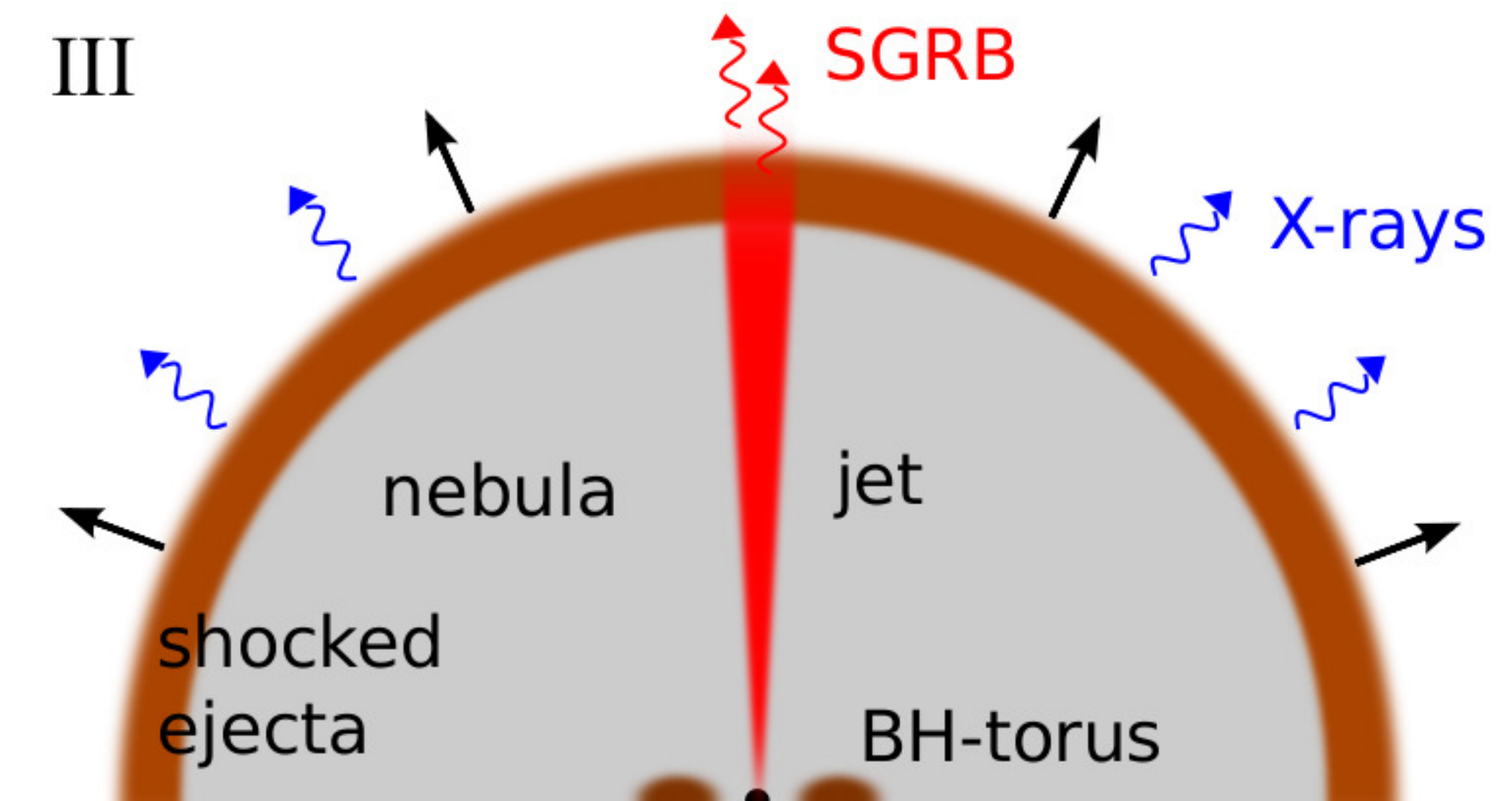}
\caption{The three phases of the ``time-reversal'' phenomenology (from \cite{Ciolfi2015}): 
(I) a baryon-loaded and highly isotropic wind is ejected by the newly-born differentially rotating SMNS; (II) spin-down radiation emitted by the cooled-down and uniformly rotating NS inflates a photon-pair plasma nebula that drives a shock through the ejecta; (III) shortly after the NS has collapsed to a BH, a relativistic jet drills through the nebula and the ejecta shell and produces the prompt SGRB, while radiation powered by the SMNS spin-down diffuses outward on a much longer timescale.}
\label{fig:phenomenology} 
\end{figure*} 

\section{Time-reversal phenomenology}\label{phenomenology}

\noindent The scenario assumes that a supramassive NS (SMNS) is formed as 
the result of a BNS merger\footnote{A long-lived NS can 
only be formed from BNS progenitors and thus the scenario excludes 
NS-BH binary mergers as the origin of SGRBs in the vast majority of observed events (i.e., in those that 
exhibit a long-lasting X-ray afterglow).}. A SMNS has a mass above 
the maximum mass for nonrotating configurations, but below the 
maximum mass for uniformly rotating configurations. 
As a consequence, the star can be supported by uniform rotation for 
a long time ($\sim$spin-down timescale) before eventually collapsing 
to a BH. 
The formation of a SMNS is favoured by the existence of NSs with a 
mass as high as $\approx 2$~M$_\odot$ \cite{Demorest2010} and by 
the BNS mass distribution \cite{Belczynski2008} 
(see, e.g., the discussion in \cite{Ciolfi2015,Siegel2015}).

The basic phenomenology of our scenario consists of three phases, which are illustrated in 
Figure~\ref{fig:phenomenology}.
During phase I, the newly-born SMNS is characterized by strong
differential rotation, magnetic fields are amplified via magnetic winding 
(and possibly other mechanisms, such as the magnetorotational instability 
\cite{Siegel2013}) and induce substantial mass ejection in the 
form of a mildly relativistic and highly isotropic, baryon-loaded wind 
\cite{Siegel2014a,Siegel2015}. Other mechanisms such as neutrino-induced outflows 
can also contribute to mass ejection. 
Within a timescale of $\lesssim 1$~s differential rotation is gradually being
removed, baryon pollution in the NS surrounding decreases and the star 
settles down to uniform rotation. At this point (phase II), the NS starts to emit 
spin-down radiation as an ordinary pulsar, inflating 
a photon-pair plasma nebula behind the expanding optically thick ejecta. 
The high photon pressure of the nebula drives a shock across the ejecta, which rapidly 
sweeps up all the material into a thin shell, in which thermal and kinetic 
energy is deposited.

After a long time (of the order of the spin-down timescale) the NS finally collapses to a BH (phase III).
The resulting BH-torus system provides the conditions to launch a relativistic jet, which can easily drill through the nebula and 
the ejecta shell, ultimately producing the prompt SGRB emission. 
The energy emitted by the NS via spin-down radiation up to the time 
of collapse diffuses outwards on much longer 
timescales, due to the high optical depth of the nebula and the ejecta. 
As a result, the associated X-ray signal will be observed for a long 
time \textit{after} the SGRB itself, as an ``afterglow'', even though the 
energy powering this emission was radiated away from the star 
\textit{before} the collapse (``time reversal'').

As a fundamental test for the time-reversal scenario, the diffusion timescale for spin-down radiation through the nebula and the ejecta immediately before the collapse has been estimated in \cite{Ciolfi2015}.
Spanning a wide range of physical parameters, the associated maximum
delay of X-rays from the system has been found generally compatible with the observed 
X-ray afterglow durations.

\section{Model predictions and supporting evidence}\label{evidence}

\noindent Besides being compatible with present observational evidence, the time-reversal phenomenology also provides very specific predictions that can be tested with future observations. 
According to the scenario, only part of the long-lasting X-ray signal should be observable after the prompt SGRB, and thus appear as an ``afterglow''. The entire signal should rather consist of long-lasting X-ray emission intercepted by the SGRB, with no apparent causal connection between the two types of emission; the time at which the burst emerges from the X-ray signal only depends on the delay associated with the radiation powered by the SMNS just before collapsing into a BH. Consequently, the observation of afterglow-like X-ray emission also prior to the SGRB would represent a strong indication in favour of the time-reversal scenario.
The observation of long-lasting X-ray emission without any SGRB counterpart would also provide support to the model. The SGRB is expected to show  a certain degree of collimation, as opposed to the isotropy of the X-ray signal predicted in our phenomenology. Therefore, such ``orphan'' events, in which the burst is beamed away from the observer, are expected to occur quite frequently. In both cases, detecting the X-ray signal without the trigger from a SGRB observation represents a challenge for present detectors, but it might become feasible in the near future. In particular, prospects for detection will improve with the combined observation of GW signals (see below). 

The peak amplitude of GWs associated with the inspiral of a BNS corresponds to the time of merger. Current GW searches with ground-based detectors usually employ the standard assumption that, if a SGRB is observed, the time of merger and thus the peak GW signal should occur within a time window of at most a few seconds from the burst. In our scenario, the SGRB occurs much later than the merger and the separation between the peak GW signal and the burst corresponds to the lifetime of the SMNS. This implies a very different strategy to maximize the chances of GW detection. If the two signals are observed with a large separation that is compatible with SMNS spin-down timescales (typically $\sim 10^2-10^4$~s \cite{Ciolfi2015}), this would provide a ``smoking gun'' evidence in favour of the time-reversal scenario. Moreover, with such large time separations a GW detection might serve as an ideal trigger for the observation of a SGRB and/or an associated long-lasting X-ray signal, provided that a sufficiently accurate estimate of the source's sky location is available (which requires multiple GW detectors).

\section{Constraints on neutron star properties}\label{constraints}

\noindent The time reversal scenario provides an opportunity to place stringent constraints on the internal structure of NSs. If confirmed, the model implies that the metastable object formed as the result of the BNS merger is a SMNS in most of the observed events.
For a given equation of state (EOS) describing the NS internal structure, the range of masses that defines a supramassive star is relatively narrow. Therefore, a reliable estimate of the NS mass could be used to exclude most EOS.

\begin{figure}[t]
\centering 
\includegraphics[angle=0,width=0.46\textwidth]{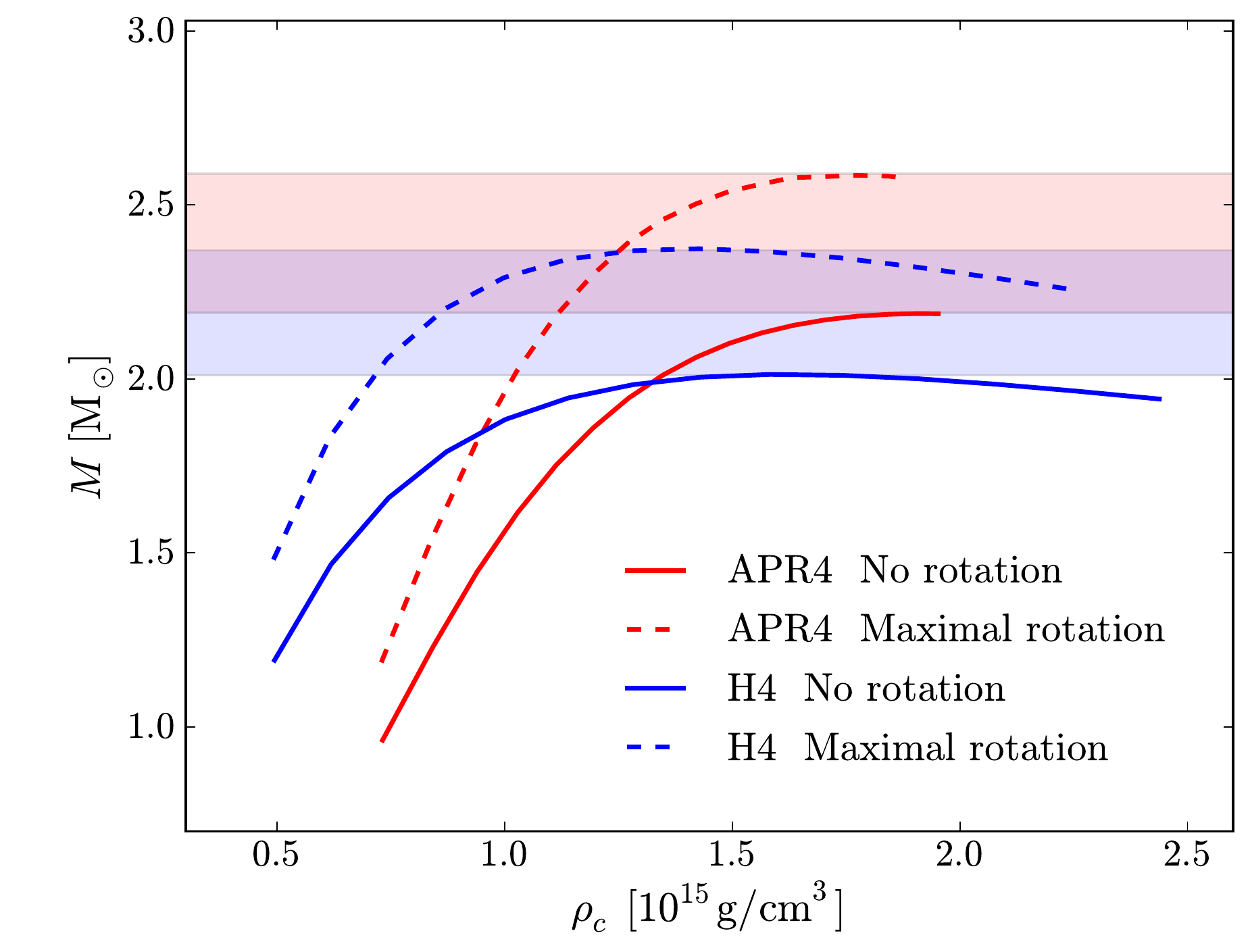}
\caption{Gravitational mass of a NS as a function of its central rest-mass density for two different EOS, APR4 (red) and H4 (blue). The continuous and dashed lines refer to the sequence of nonrotating and maximally rotating configurations, respectively. The shaded regions indicate the SMNS range for the two EOS.}
\label{fig:eos} 
\end{figure}

In Figure~\ref{fig:eos} we show the gravitational mass of a NS as a function of the central rest-mass density for two different EOS, APR4 and H4 \cite{Akmal1998}. The continuous and dashed lines refer to a sequence of nonrotating and maximally (uniformly) rotating configurations, respectively. The NS is supramassive if the mass is between the maxima of these two profiles. Above the maximum of the upper curve, the NS is hypermassive and it will collapse to a BH as soon as differential rotation is removed (typically within $\lesssim 1$~s), while below the maximum of the lower curve the star is stable and it will never collapse even if all the rotational energy is removed. 
As an illustrative example, suppose that we have an estimate for the NS gravitational mass of $\approx 2.34$~M$_\odot$. According to the simple formula $M\simeq 0.9(M_1+M_2-0.1$~M$_\odot)$ proposed in \cite{Belczynski2008}, which relates the masses $M_1, M_2$ of the two NSs of the binary system to the mass $M$ of the merged object, this mass would correspond to a 1.3$-$1.4~M$_\odot$ BNS. Both APR4 and H4 would be compatible with the requirement of a supramassive star for this mass estimate. For a final mass of $\approx 2.43$~M$_\odot$ (corresponding to a 1.4$-$1.4~M$_\odot$ BNS), only APR4 would be compatible while H4 would be excluded. Finally, for $M\approx 2.61$~M$_\odot$ (corresponding to a 1.5$-$1.5~M$_\odot$ BNS), both EOS would be excluded.

Combining a number of joint electromagnetic and GW observations will allow us to significantly reduce the set of possible EOS. The basic requirement to achieve this goal is to extract an estimate for $M$. GW observations can provide a measure of the chirp mass of the binary, which can be translated into mass estimates of the individual NSs assuming a certain mass ratio. The SMNS mass can then be inferred from a formula like the one mentioned above. Assuming a mass ratio of 0.8$-$1, for instance, the total mass $M$ should be known to an accuracy of $\sim 10$\%. 

Furthermore, the observation of both the SGRB and the peak GW signal produced at the merger would provide a very accurate measure of the SMNS lifetime\footnote{For a typical spin-down timescale of $\sim 10^3$~s, the $\sim$s precision in determining the time of merger and the time of collapse to a BH would result in a $\sim 0.1$\% accuracy.}. This additional information can be used to further constrain NS properties.
Given a reliable NS mass estimate (e.g., as discussed above) and assuming an EOS that is compatible with the supramassive requirement, one can directly estimate the spin period of the NS at the time of collapse. If the loss of rotational energy is due to dipole spin-down radiation, for each initial spin period there is only one magnetic field strength $B_\mathrm{p}$ (surface dipolar component) compatible with the measured NS lifetime. Hence, restricting to a reasonable range of initial spin periods (e.g., $\sim 0.5-2$~ms) would limit $B_\mathrm{p}$ to a relatively narrow range. If an independent measurement of the initial spin period is available (e.g, via GW observations) one would obtain a precise estimate of $B_\mathrm{p}$. Alternatively, an estimate of the magnetic field strength (possibly from the prompt SGRB and/or X-ray afterglow luminosities) would yield the initial spin period. 

\section{Concluding remarks}\label{conclusion}

\noindent The time-reversal scenario provides a possible solution to the present dichotomy posed by the observation of long-lasting X-ray afterglows in a large fraction of SGRB events, and thus represents an attractive alternative to current SGRB models. A first test concerning the estimated delays that affect the emission powered by the SMNS spin-down shows a broad compatibility with the observed X-ray afterglow durations.
Moreover, the scenario is characterized by very specific predictions that can be tested with future observations.
These include the presence of afterglow-like X-ray emission prior to the SGRB or the possibility of observing an ``orphan'' event, in which afterglow-like X-ray emission is not accompanied by a SGRB. In addition, the peak of GW emission associated with the BNS merger is expected to occur much earlier than the SGRB, with the two signals being separated by the lifetime of the SMNS. This might provide a very precise measure of the lifetime and allow us to employ GW detections to trigger electromagnetic observations of SGRBs and/or of the associated long-lasting X-ray emission. If future observational evidence supports the model, it will provide a solid astrophysical framework to understand the physical mechanisms that generate SGRBs and it will allow us to place important constrains on NS properties.

Further investigation is necessary in order to confirm the viability of the time reversal phenomenology.  
More accurate predictions are needed for the first phase of the SMNS evolution, in which the star is differentially rotating and baryon-loaded winds are produced. This requires accurate general-relativistic magnetohydrodynamic simulations of BNS mergers that lead to the formation of a SMNS. Moreover, the dynamics of the system on length and time scales characterizing the second and third phase of the scenario need to be studied in more detail, and predictions need to be compared with X-ray afterglow observations. 
Finally, a potential challenge for the scenario is posed by the formation of a jet from the collapse of a uniformly rotating magnetized SMNS. While some of the necessary conditions to launch a relativistic jet (e.g., the presence of a massive accretion torus) have been found in previous numerical studies for the case in which a BH-torus system is formed at merger, analogous evidence is missing in the case of the delayed collapse envisaged by the time-reversal scenario.



\end{document}